\documentclass{article}
\usepackage{cite}
\usepackage{graphicx}
\usepackage{dcolumn}

\begin{document}

\date{}
\title{The harmonic oscillator in a space with a screw dislocation}
\author{Paolo Amore\thanks{%
e--mail: paolo.amore@gmail.com} \\
%EndAName
Facultad de Ciencias, CUICBAS, Universidad de Colima,\\
Bernal D\'{i}az del Castillo 340, Colima, Colima,Mexico \\
and \\
Francisco M. Fern\'andez\thanks{%
e--mail: fernande@quimica.unlp.edu.ar} \\
INIFTA (CONICET), Divisi\'{o}n Qu\'{i}mica Te\'{o}rica,\\
Blvd. 113 y 64 (S/N), Sucursal 4, Casilla de Correo 16,\\
1900 La Plata, Argentina}
\maketitle

\begin{abstract}
We obtain the eigenvalues of the harmonic oscillator in a space with a screw
dislocation. By means of a suitable nonorthogonal basis set with variational
parameters we obtain sufficiently accurate eigenvalues for an arbitrary
range of values of the space-deformation parameter. The energies exhibit a
rich structure of avoided crossings in terms of such model parameter.
\end{abstract}

\section{Introduction}

\label{sec:intro}

Space dislocations have been useful for the description of a variety of
physical phenomena. Among such applications we mention an analysis of the
influence of frozen-in topological defects in a crystal on the
long-wavelength quantum states of a particle\cite{BST98}, a study of
electrons moving in a magnetic field in the presence of a screw dislocation%
\cite{FM99}, the scattering of electrons on a screw dislocation\cite{BST99},
an investigation of the quantum scattering of an electron by a screw
dislocation with an internal magnetic flux\cite{FBM01}, a geometric model
for the explanation of the origin of the observed shallow levels in
semiconductors threaded by a dislocation density\cite{BM12}, the influence
of the Aharonov-Casher effect on the Dirac oscillator in three different
scenarios of general relativity\cite{BF13}, an investigation of torsion and
noninertial effects on a spin-1/2 quantum particle in the nonrelativistic
limit of the Dirac equation\cite{B14}, a study of ac electronic transport in
semiconductor crystals with a screw dislocation\cite{TS14}, a
two-dimensional electron gas on a cylindrical surface with a screw
dislocation\cite{FS15}, a study of spin currents induced by topological
screw dislocation and cosmic dispiration\cite{WMLF15}, an analysis of a
relativistic scalar particle with a position-dependent mass in a spacetime
with a space-like dislocation\cite{VB16}, a study of the influence of a
screw dislocation on the energy levels and the wavefunctions of an electron
confined in a two-dimensional pseudoharmonic quantum dot under the influence
of an external magnetic field and an Aharonov-Bohm field\cite{FRAS16} and
the effect of a screw dislocation on an anharmonic oscillator\cite{B17}.

The present paper is motivated by those of Filgueiras et al\cite{FRAS16} and
Bakke\cite{B17} who solve the Schr\"{o}dinger equation with a screw
dislocation. In the former case the authors choose a deformed potential $%
V_{d}(\rho )$ and a scalar pseudoharmonic interaction $V_{\mathrm{conf}%
}(\rho )$, $\rho ^{2}=x^{2}+y^{2}$, so that the motion of the electron along
the $z$-axis is free. Under these conditions the resulting eigenvalue
equation is separable and exactly solvable. Later the authors consider that
the electrons are confined by infinite walls at $z=0$ and $z=d$ and claim
that the eigenvalue equation is still separable. In the latter case the
author chooses a potential $V(\rho )$ so that the motion of the particle is
unbounded along the $z$ direction and the spectrum continuous. Here we
choose one of the simplest confining potentials, the three-dimensional
harmonic oscillator, and obtain approximate eigenvalues of the resulting
nonseparable deformed Schr\"{o}dinger equation. In section~\ref{sec:model}
we develop the main equations for the model, in section~\ref{sec:variational}
we first obtain approximate results by means of a simple variational ansatz
that later use as the starting point of a more accurate Rayleigh-Ritz
variational calculation. In this section we show results for different
quantum numbers in a range of values of the dislocation parameter. Finally,
in section\ref{sec:conclusions} we summarize the main results and draw
conclusions.

\section{The model}

\label{sec:model}

Some kind of topological defects are described by means of the line element
\begin{equation}
ds^{2}=g_{ij}dy^{i}dy^{j},  \label{eq:ds2_general}
\end{equation}
where $g_{ij}$ are the elements of the metric tensor, $\{y^{i}\}$ is a
suitable set of curvilinear coordinates and summation on repeated indices is
assumed. The Laplacian in such a space is given by
\begin{equation}
\nabla ^{2}=\frac{1}{\sqrt{|\mathbf{g}|}}\partial _{i}\sqrt{|\mathbf{g}|}%
g^{ij}\partial _{j},  \label{eq:Laplacian_general}
\end{equation}
where $|\mathbf{g}|$ is the determinant of the matrix $\mathbf{g}=\left(
g_{ij}\right) $, $\partial _{i}=\frac{\partial }{\partial y^{i}}$ and $%
g^{ij}g_{jk}=\delta _{k}^{i}$. Valanis and Panoskaltsis\cite{VP05} derive
expressions for a wide variety of deformations in a material body.

The Hamiltonian operator for a particle of mass $m$ moving in such a space
under the effect of a potential-energy function $V(\mathbf{r})$ is
\begin{equation}
H=-\frac{\hbar ^{2}}{2m}\nabla ^{2}+V(\mathbf{r}).  \label{eq:H}
\end{equation}

In order to solve the Schr\"{o}dinger equation for $H$ it is convenient to
choose a convenient set of units. We choose a set of dimensionless
coordinates $\mathbf{r}^{\prime }=\mathbf{r}/L$, where $L$ is an arbitrary
length, and rewrite the Hamiltonian operator in dimensionless form as
\begin{equation}
\frac{2mL^{2}}{\hbar ^{2}}H=\nabla ^{\prime 2}+v(\mathbf{r}^{\prime }),\;v(%
\mathbf{r}^{\prime })=\frac{2mL^{2}}{\hbar ^{2}}V(L\mathbf{r}^{\prime }),
\label{eq:H_dim}
\end{equation}
where $\nabla ^{\prime 2}=L^{2}\nabla ^{2}$

Following Filgueiras et al\cite{FRAS16} and Bake\cite{B17} we choose the
screw dislocation
\begin{equation}
ds^{2}=d\rho ^{2}+\rho ^{2}d\phi ^{2}+\left( dz+\eta d\phi \right) ^{2},
\label{eq:ds2_screw}
\end{equation}
where $\rho =\sqrt{x^{2}+y^{2}}$, $\phi =\arctan \left( y/x\right) $ and $%
\eta $ characterizes the torsion field (dislocation). In this case the
Laplacian is
\begin{equation}
\nabla ^{2}=\frac{1}{\rho }\partial _{\rho }\rho \partial _{\rho }+\partial
_{z}^{2}+\frac{1}{\rho ^{2}}\left( \partial _{\phi }-\eta \partial
_{z}\right) ^{2}.  \label{eq:Laplacian_screw}
\end{equation}
If we choose a harmonic interaction
\begin{equation}
V(r)=\frac{k}{2}r^{2}=\frac{k}{2}\left( \rho ^{2}+z^{2}\right) ,
\label{eq:V_HO}
\end{equation}
and the length unit $L=\sqrt{\frac{\hbar }{m\omega }}$, $\omega =\sqrt{\frac{%
k}{m}}$, the dimensionless Schr\"{o}dinger equation becomes
\begin{eqnarray}
&&\left[ -\frac{1}{\rho }\partial _{\rho }\rho \partial _{\rho }-\partial
_{z}^{2}-\frac{1}{\rho ^{2}}\left( \partial _{\phi }-\lambda \partial
_{z}\right) ^{2}+\rho ^{2}+z^{2}\right] \psi =\mathcal{E}\psi ,\;  \nonumber
\\
&&\mathcal{E}=\frac{2mL^{2}}{\hbar ^{2}}E,\;\lambda =\frac{\eta }{L},
\label{eq:Schro_HO_dim}
\end{eqnarray}
where we have omitted the primes in $\rho ^{\prime }$ and $z^{\prime }$.

If we write
\begin{equation}
\psi \left( \rho ,z,\phi \right) =F(\rho ,z)e^{im\phi },\;m=0,\pm 1,\pm
2,\ldots   \label{eq:Psi}
\end{equation}
then we are left with an equation for $F(\rho ,z)$
\begin{equation}
\left[ -\frac{1}{\rho }\partial _{\rho }\rho \partial _{\rho }-\left( 1+%
\frac{\lambda ^{2}}{\rho ^{2}}\right) \partial _{z}^{2}+\frac{2im\lambda }{%
\rho ^{2}}\partial _{z}+\frac{m^{2}}{\rho ^{2}}+\rho ^{2}+z^{2}\right]
F(\rho ,z)=\mathcal{E}F(\rho ,z).  \label{eq:Schro_dim_2}
\end{equation}
Note that this equation is invariant under the transformations $%
(m\rightarrow -m,$ $z\rightarrow -z)$ and $(\lambda \rightarrow -\lambda ,$ $%
z\rightarrow -z)$, therefore the eigenvalues depend on $m^{2}$ and $\lambda
^{2}$.

When $\lambda =0$ the equation is fully separable $F_{nkm}(\rho
,z)=f_{nm}(\rho )g_{k}(z)$ and the energy eigenvalues $\mathcal{E}_{nkm}(0)$
are $g_{nkm}$-fold degenerate, where
\begin{eqnarray}
\mathcal{E}_{nkm}(0) &=&4n+2k+2|m|+3,\;n,k=0,1,\ldots ,  \nonumber \\
g_{nkm} &=&\frac{1}{2}\left( 4n+2k+2|m|\right) \left( 4n+2k++2|m|+1\right) .
\label{eq:Enkm(0)}
\end{eqnarray}
Note that the dislocation removes almost all the degeneracy leaving only
that coming from the $z$-component of the angular momentum. In what follows
we use the quantum numbers $n$, $k$ and $|m|$ to label the eigenvalues of
the nonseparable Schr\"{o}dinger equation for $\lambda =0$. Obviously, only $%
m$ is a good quantum number.

\section{Variational approach}

\label{sec:variational}

In this section we try to obtain approximate eigenvalues to equation (\ref
{eq:Schro_dim_2}). Our starting point is the simple (unnormalized)
variational function for the states with $n=k=0$:
\begin{equation}
\varphi (\rho ,z)=\rho ^{s}\exp \left( -\rho ^{2}/2-bz^{2}\right) ,
\label{eq:psi_var_simple}
\end{equation}
where $s$ and $b$ are variational parameters. A straightforward calculation
shows that the minimum of
\begin{equation}
W(s,b)=\frac{\left\langle \varphi \right| H\left| \varphi \right\rangle }{%
\left\langle \varphi \right. \left| \varphi \right\rangle },
\label{eq:W_var}
\end{equation}
is given by
\begin{equation}
s=\sqrt{b\lambda ^{2}+m^{2}},\;\left( 4b^{2}-1\right) \sqrt{b\lambda
^{2}+m^{2}}+4b^{2}\lambda ^{2}=0,  \label{eq:s_simple}
\end{equation}
and the upper bound to the lowest energy for a given value of $|m|$ is
\begin{equation}
W=\frac{8b\sqrt{b\lambda ^{2}+m^{2}}+4b^{2}+8b+1}{4b}.
\label{eq:W_opt__simple}
\end{equation}

In order to improve the accuracy of the results and obtain the excited-state
energies we apply the Rayleigh-Ritz variational method with the
nonorthogonal basis set
\begin{equation}
\varphi (\rho ,z)=\sum_{i=0}^{M-1}\sum_{j=0}^{N-1}c_{ij}\rho ^{i+s}z^{j}\exp
\left( -\rho ^{2}/2-bz^{2}\right) ,  \label{eq:psi_RR}
\end{equation}
where $s$ is given by equation (\ref{eq:s_simple}). In this case we obtain
the minimum of the variational integral (\ref{eq:W_var}) with respect to the
linear parameters $c_{ij}$ and to the nonlinear one $b$.

\begin{figure}[tbp]
\begin{center}
\includegraphics[width=6cm]{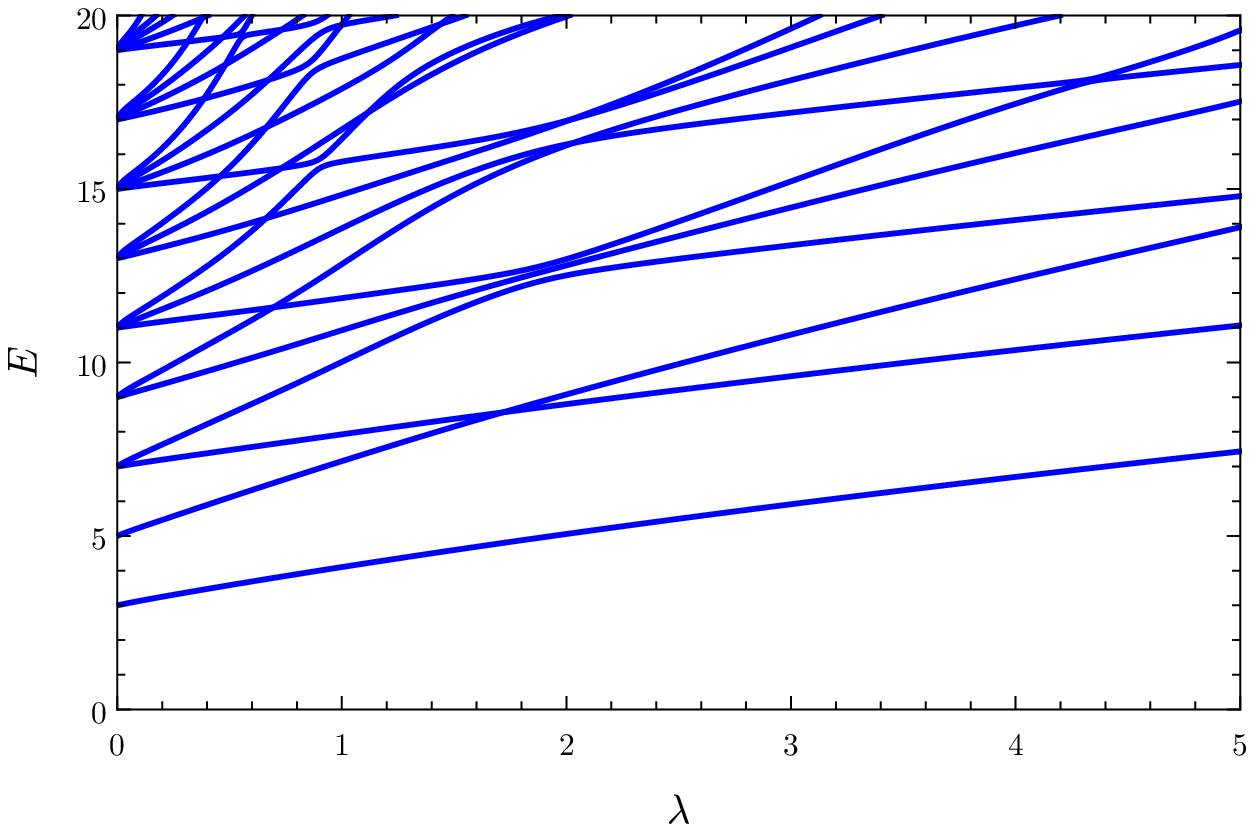} \includegraphics[width=6cm]{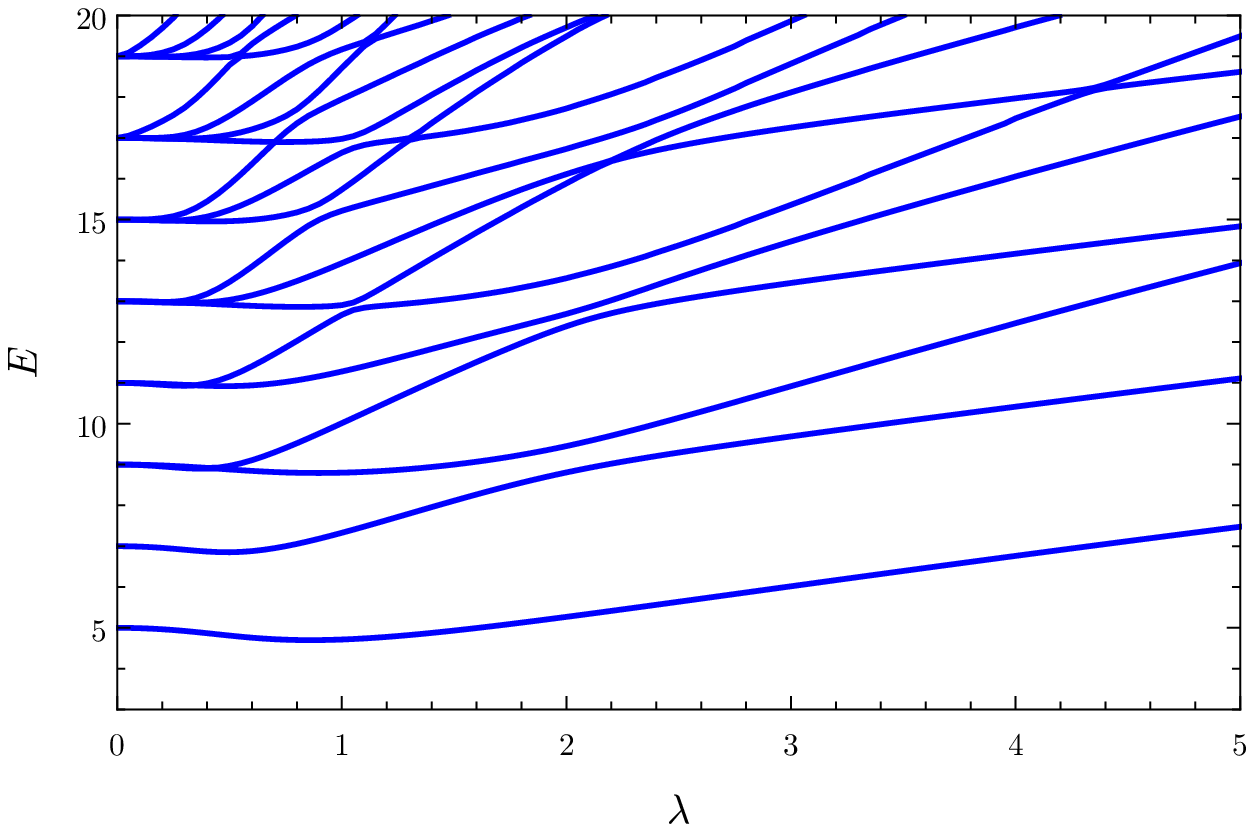} %
\includegraphics[width=6cm]{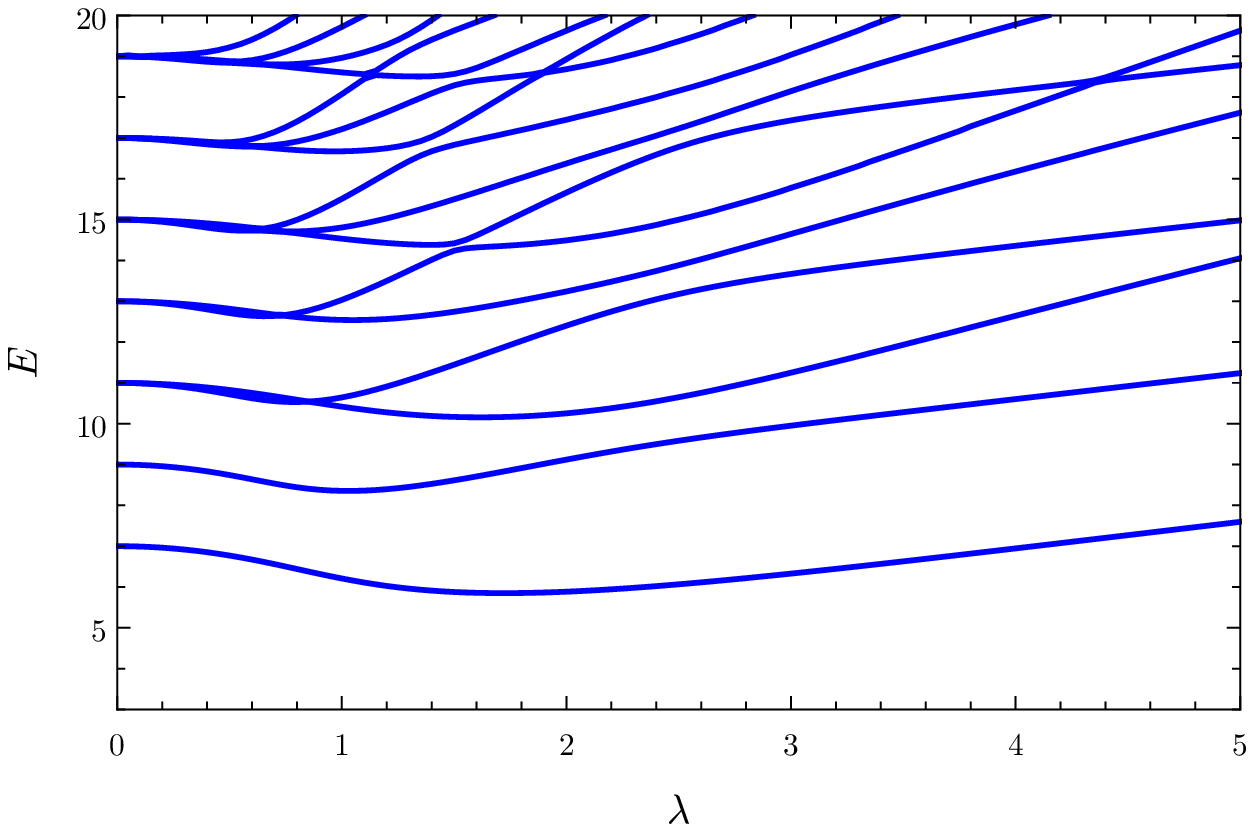}
\end{center}
\caption{Lowest eigenvalues with $|m|=0,1,2$}
\label{fig:ene}
\end{figure}

Figure~\ref{fig:ene} shows the lowest eigenvalues $\mathcal{E}_{nkm}(\lambda
)$ for $|m|=0,1,2$ in the interval $0\leq \lambda \leq 5$. We appreciate
that the energies exhibit a rich structure of avoided crossings. Such
avoided crossings take place because eigenvalues with the same quantum
number $m$ do not cross. The arguments for the noncrossing rule are similar
to those for the states with the same symmetry already discussed by several
authors\cite{T37, RNBB72,F14,F15}.

\begin{figure}[tbp]
\begin{center}
\includegraphics[width=6cm]{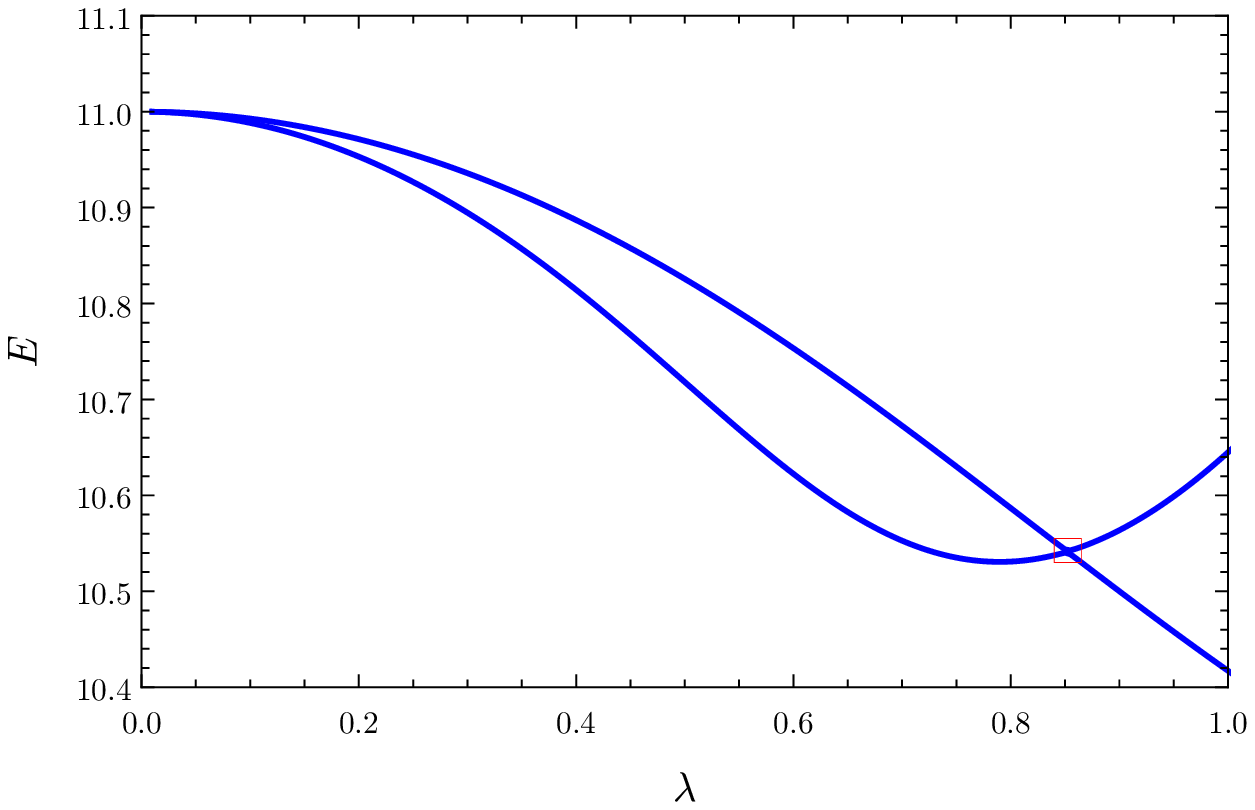} \includegraphics[width=6cm]{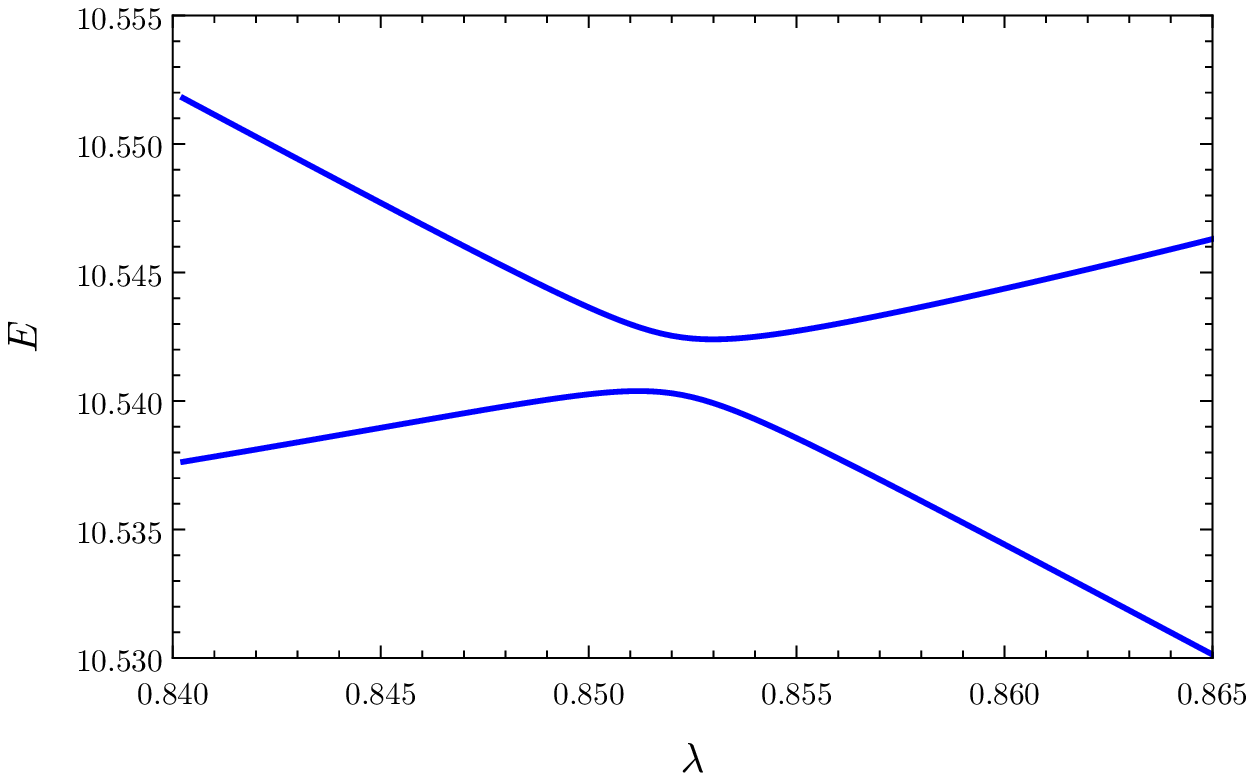} %
\includegraphics[width=6cm]{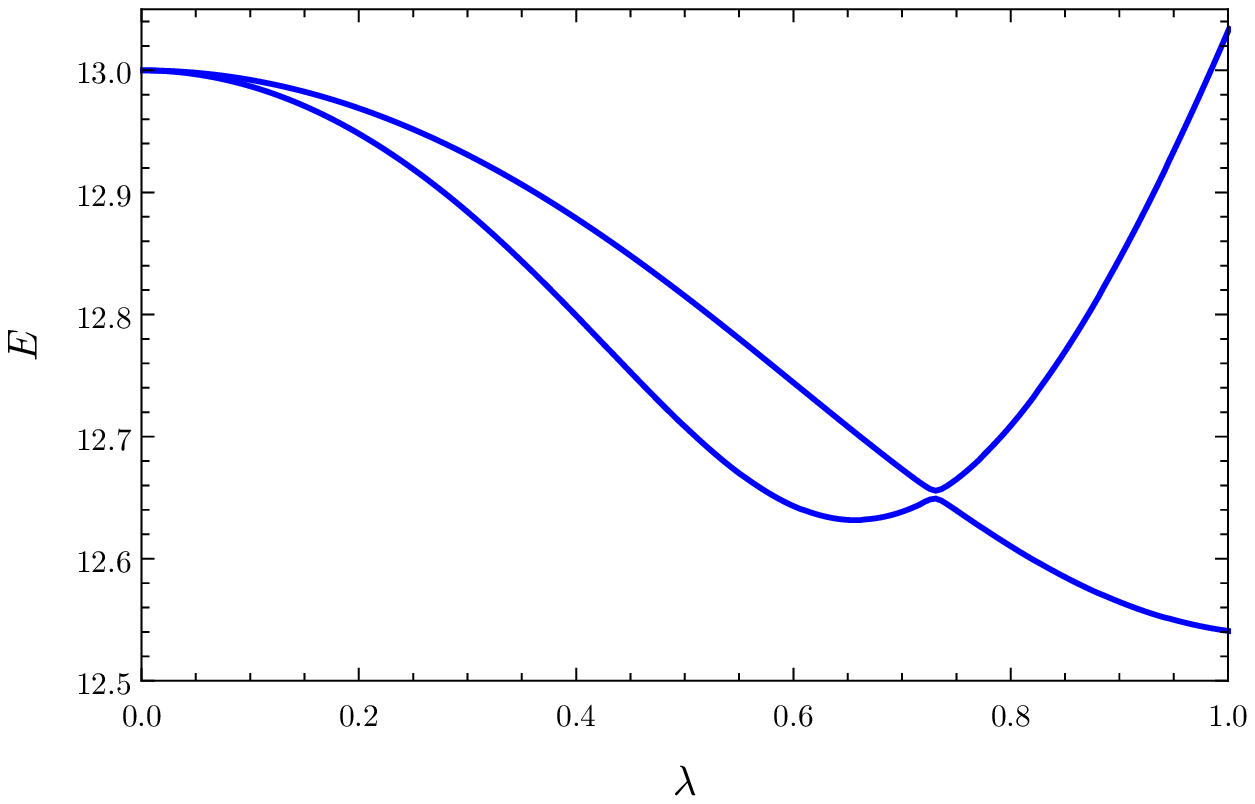} \includegraphics[width=6cm]{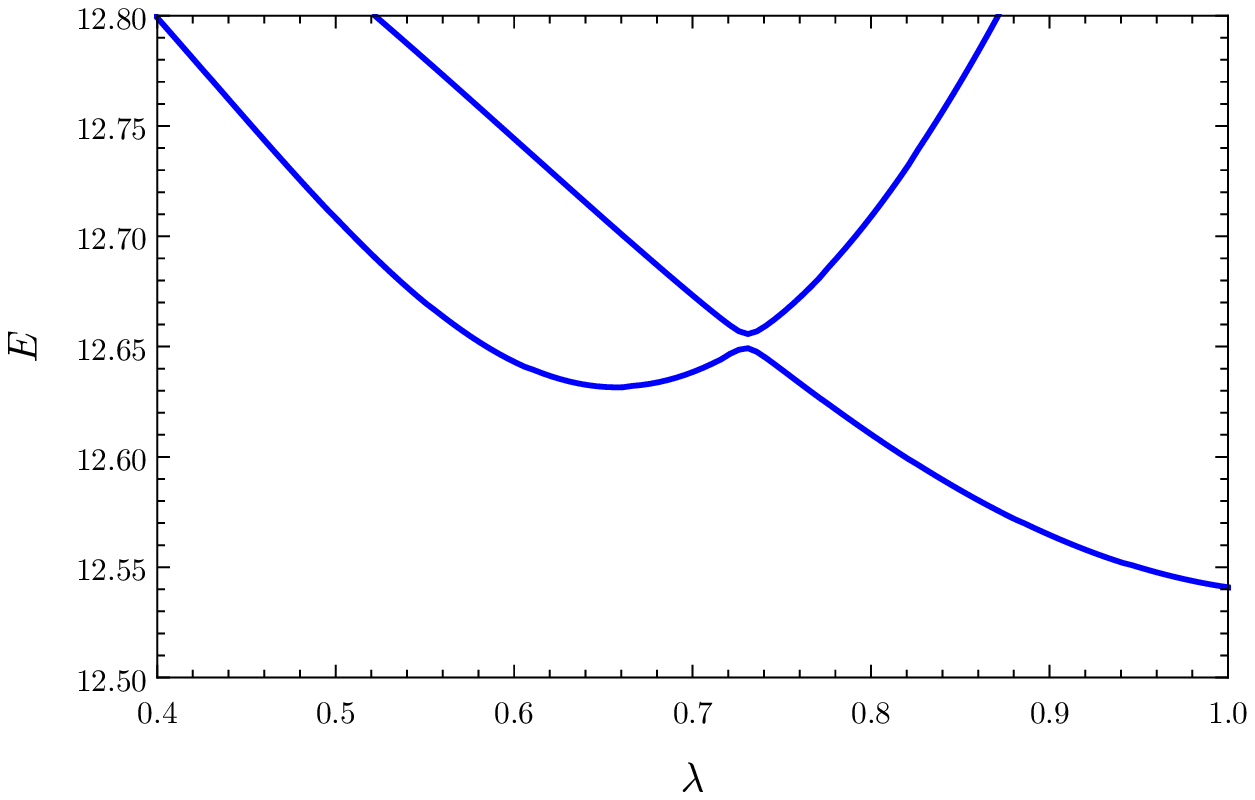} %
\includegraphics[width=6cm]{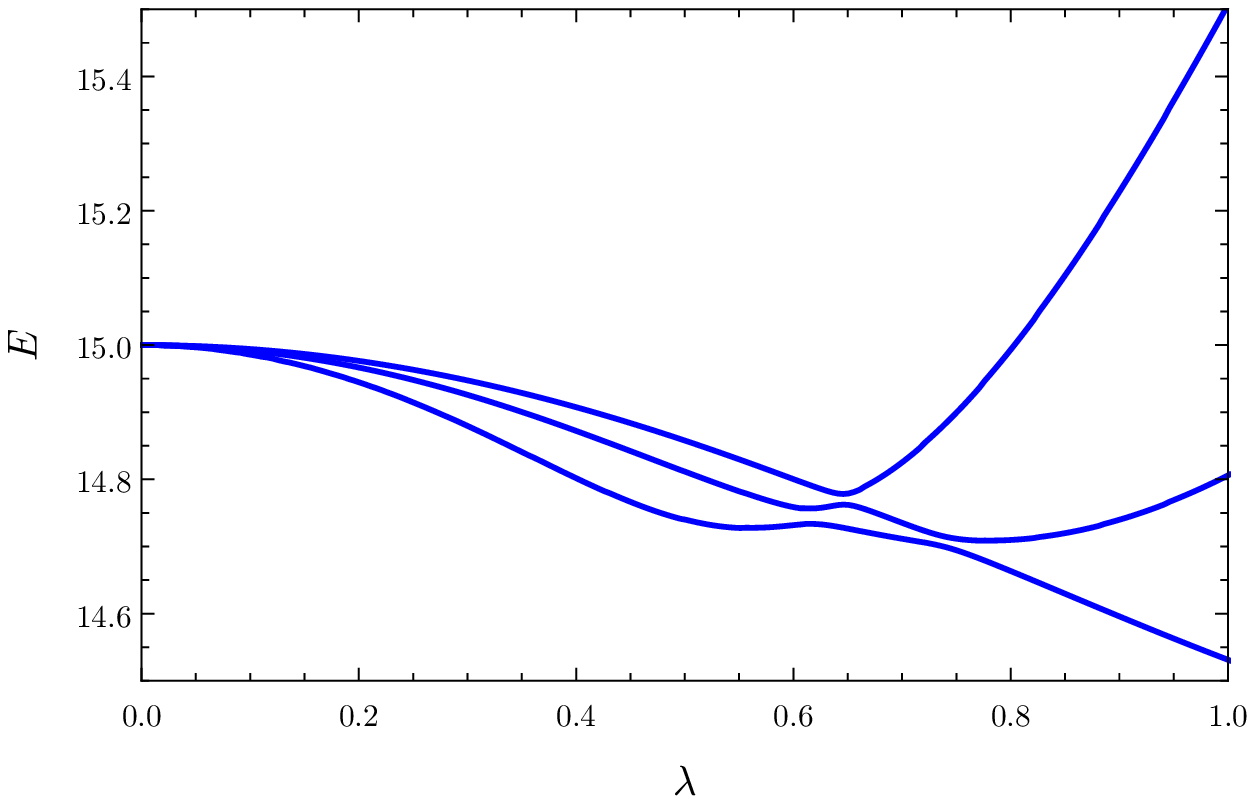}
\end{center}
\caption{Some multiplets of eigenvalues with $|m|=2$}
\label{fig:E2DET}
\end{figure}

Some of the avoided crossings shown in those figures may appear to be actual
crossings because the eigenvalues approach each other quite closely. In
order to illustrate this point, Figure~\ref{fig:E2DET} shows the avoided
crossings between eigenvalues in the multiplets $\left( \mathcal{E}_{102},%
\mathcal{E}_{022}\right) $, $\left( \mathcal{E}_{112},\mathcal{E}%
_{032}\right) $ and $\left( \mathcal{E}_{202},\mathcal{E}_{122},\mathcal{E}%
_{042}\right) $ in somewhat more detail.

The first pair of subfigures in Figure~\ref{fig:E2DET} show the eigenvalues $%
\left( \mathcal{E}_{102},\mathcal{E}_{022}\right) $ stemming from $\mathcal{E}%
(0)=11$. The right subfigure is just an enlargement of that part of the left
one corresponding to the avoided crossing. The second pair of subfigures
illustrate the same feature for the pair of eigenvalues $\left( \mathcal{E}%
_{112},\mathcal{E}_{032}\right) $ stemming from
$\mathcal{E}(0)=13$. The last subfigure shows the triplet of
states stemming from $\mathcal{E}(0)=15$.

What is interesting in this model is that the eigenvalues of a given
multiplet with a fixed value of $|m|$ exhibit avoided crossings among
themselves for sufficiently small values of $\lambda $ (for example $%
0<\lambda <1$). For greater values of $\lambda $ there are avoided crossings
between eigenvalues of different multiplets as shown in Figure~\ref{fig:ene}.

\section{Conclusions}

\label{sec:conclusions}

As stated in the introduction this paper is motivated by those of Filgueiras
et al\cite{FRAS16} and Bakke\cite{B17}. The main difference is that the
models chosen by those authors are separable and do not reveal the
possibility of avoided crossings. In addition to this, in those models the
motion of the particle is unbounded in the $z$ direction and the
corresponding spectra are continuous. Filgueiras et al\cite{FRAS16}
attempted to confine the electron in the $z$ axis by means of a square-well
potential but they did not take into account the correct boundary
conditions. Their wavefunctions satisfy $\psi (\rho ,\varphi ,d)=e^{i\ell
\pi }\psi (\rho ,\varphi ,0)$, where $\psi (\rho ,\varphi ,0)\neq 0$, and,
consequently, their results apply to a problem with boundary conditions $%
\psi (\rho ,\varphi ,d)=\pm \psi (\rho ,\varphi ,0)$ and not to the box
confinement. If the walls of the square well are located at $z=0$ and $z=d$
then the Schr\"{o}dinger equation with the correct boundary conditions $\psi
(\rho ,\varphi ,0)=\psi (\rho ,\varphi ,d)=0$ is not separable.

In this paper we have chosen a harmonic-oscillator potential in
order to investigate the effect of space dislocation on the
spectrum. The reason for this choice is that this potential is
well known and facilitates the calculation of the matrix elements
necessary for the application of the variational method. Besides,
the Schr\"{o}dinger equation is exactly solvable when there is no
dislocation and we can therefore easily appreciate how the
dislocation breaks the degeneracy. Although this model may not
have a clear physical application it nevertheless shows what one
may expect from a realistic model with a confining potential and
screw dislocation. Most probably such model will also exhibit some
kind of structure of avoided crossings. The screw dislocation not
only breaks the degeneracy but also gives rise to avoided
crossings between energy levels.

\end{document}